# Promotional Language and the Adoption of Innovative Ideas in Science

Hao Peng[1,2*], Huilian Sophie Qiu[1,2*], Henrik Barslund Fosse[3], and Brian Uzzi[1,2**]

[1]Kellogg School of Management, Northwestern University, Evanston, IL 60208
[2]Northwestern Institute on Complex Systems, Evanston, IL 60208
[3]Novo Nordisk Foundation, Hellerup, Denmark 2900

* Equal Contribution; ** Corresponding Author

## Abstract

How are the merits of innovative ideas communicated in science? Here we conduct semantic analyses of grant application success with a focus on scientific promotional language, which has been growing in frequency in many contexts and purportedly may help to convey an innovative idea's originality and significance. Our analysis attempts to surmount limitations of prior studies by examining the full text of tens of thousands of both funded and unfunded grants from three leading public and private funding agencies: the NIH, the NSF, and the Novo Nordisk Foundation, one of the world's largest private science funding foundations. We find a robust association between promotional language and the support and adoption of innovative ideas by funders and other scientists. First, the percentage of promotional language in a grant proposal is associated with up to a doubling of the grant's probability of being funded. Second, a grant's promotional language reflects its intrinsic level of innovativeness. Third, the percentage of promotional language predicts the expected citation and productivity impact of publications that are supported by funded grants. Lastly, a computer-assisted experiment that manipulates the promotional language in our data demonstrates how promotional language can communicate the merit of ideas through cognitive activation. With the incidence of promotional language in science steeply rising, and the pivotal role of grants in converting promising and aspirational ideas into solutions, our analysis provides empirical evidence that promotional language is associated with effectively communicating the merits of innovative scientific ideas.

## Significance Statement

Using three longitudinal samples of funded and unfunded grant applications from three of the world's largest funders—the NIH, the NSF, and the Novo Nordisk Foundation—we find that the percentage of promotional language in a grant proposal is associated with a grant's probability of being funded, estimated level of innovativeness, and predicted levels of citation impact. Further, computer experiments show that promotional words may elevate a grant's positive impressions. While scientific ideas should be evaluated by their intrinsic quality, not by their linguistic packaging, our results empirically suggest that promotional language can harmonize the substance and presentation of scientific work to effectively communicate the merits of innovative ideas.

## Introduction

Converting scientific curiosity into facts and innovation frequently depends on a research team's ability to effectively convey the merits of new ideas. This communication process is important in most areas of science, including publications, symposia, the media, and especially grant funding, which is a pivotal first step in the communication process (1-4). Besides being a gateway to resources that support research, grant writing is one of a scientist's most time-consuming efforts. Annually, scientists spend 66 of their 260 working days on funding acquisition. The coveted RO1 grant (a.k.a. the "tenure award" grant), for example typically adds up to millions of hours of researchers' and reviewers' time each year (5, 6).

Despite the importance of funding for science, knowledge about how scientists communicate the merits of their ideas to funders is only partly understood (7-10). For example, while grants are expressly designed to promote innovation, recent research indicates that reviewers commonly reject novel grants (11-13), a condition that may be linked to career setbacks (6) and to science becoming less disruptive and replicable (14, 15). Gaps also exist in the data availability in prior empirical studies which have generally lacked the full text and PI data on funded and unfunded applications that can overcome selection and model specifications biases needed for broadly representative empirical findings (6, 16-18).

A growing area of interest in science communication is the semantic analysis of large corpuses of text (19). Evidence from linguistic studies suggests that the language used in communication is associated with changes in reader's cognition of, or connections between, concepts and objects (20, 21). For example, subtle changes in the language used in college admission essays has been linked with college admission acceptance rates (22). Words in the abstracts of funded grants correlate with the size of the award (23). Evidence also suggests that promotional language may play a substantial role in communicating a grant proposal's merits by helping to convey its originality, novelty, and significance to reviewers (24, 25). A recent study showed that from 1985 to 2020, the use of promotional language steeply rose in accepted NIH grants (25). In 1985, there were approximately 6,000 promotional words per million words in funded NIH proposals. By 2020, the number had more than doubled to 13,000 per million words.

While the rise and fall in the use of certain words are themselves a natural part of the evolution of language and writing, the implications of using promotional language for communicating the quality of ideas remains unknown and debated. On the one hand, if promotional language helps to convey the merits of complex and risky ideas, it can facilitate the adoption of novel ideas in science and society (26, 27). On the other hand, if promotional language overstates findings (28), it weakens science's reputation for presenting unvarnished and reproducible facts (9, 29, 30).

To study science communication processes from the perspective of semantic analysis, we investigated the statistical relationship between a grant's promotional language and its (i) funding success, (ii) inherent innovativeness, and (iii) future impact. Empirically, our unique data surmounts some limitations of prior work. Going beyond studies of funded proposals, we analyze tens of thousands of funded and non-funded grant applications from three influential funding institutions: the NIH, the NSF, and the Novo Nordisk Foundation (NNF), one of the world's largest private sources of scientific grants, with nearly a billion dollars awarded annually. These grant datasets are further enriched with performance data on the prior grant application and publication histories of principle investigators (PIs) who submitted the grants. Lastly, a computer-assisted experiment that manipulates the promotional language in our grant data provides one possible explanation of how promotional language may communicate the merits of novel ideas.

**Data and Design**

The Novo Nordisk Foundation (NNF) data includes 13,520 grant proposals submitted between 2015 and 2022. Each proposal contains the full text of the project, program area, funding amount applied for, funding decision, and deidentified personal information, including the applicant's self-reported age, gender, publication and citation records, and grant-supported future publications. The average award size in NNF data is $600,000 (4.2 million Danish kroner), and the positive funding rate is 16.8%. The average NNF proposal length is 2,859 words.

The NIH and NSF datasets include 2,649 NIH and 561 NSF grant proposals submitted by PI faculty members at a leading research university. These deidentified data include the application year, project description, amount of funding applied for, acceptance decision, as well as the main applicant's gender, and publication and citation records. In our data, the average award sizes and funding rates are higher than the national average for NIH and NSF grants, which is expected given that our sample is from a leading research university with a medical school. The national average NIH award size is $466,000 and the funding rate is 33% across all grants. In our data, the average award size is $825,000 and the funding rate is 21.4%. For NSF, the national average award size is $361,000 and the funding rate is 25%, and in our data the average award size is $321,000 and the funding rate is 36.0%. The average NIH and NSF proposal lengths are 9,223 and 9,030 words respectively.

Our lexicon of promotional language uses a validated dictionary of 139 science-specific promotional words such as "unique," "revolutionary," or "fundamental" (See the Materials and Methods section for the full lexicon). Millar et al. (25) created this dictionary of scientific promotional language by analyzing the linguistic content of all 901,717 funded NIH grant applications from 1985 to 2020 (24, 25). To manually code and validate this lexicon, Millar et al. had two independent experts identify candidate promotional words in the NIH grants. Each candidate word was evaluated based on over 500 different instances of its use in context, and on its substitutability with a neutral synonym (Cohen's $k = 0.82$). The validation process resulted in a final set of 139 scientific promotional words used in grants. Table 1 displays examples of promotional words and their contextual usage reported by Millar et al. (25).

Adding to Millar et al.'s validation, we independently conducted three additional validation checks of the lexicon. (i) First, to externally validate the dictionary, we used the multitrait-multimethod approach (MTMM)(31). According to MTMM, if promotional words operate as expected, by helping communicate the perceived originality and significance of innovative ideas to readers (25), promotional words should correlate more strongly than their neutral synonyms with words that engender cognitive engagement. A word's valance and arousal levels have been shown to engender cognitive processing attention (32). As predicted, we found that promotional words have statistically higher average valence and arousal scores than their neutral synonyms using weighted average t-test, signrank test, or mvtest of means ($p < 0.002$ in all tests). See details in *SI Appendix*, Section I. (ii) Second, to test the internal consistency of the dictionary, we computed Cronbach's alpha for all 139 promotional words based on their percentage frequencies in our datasets, which are at acceptable levels for each dataset (Cronbach's alpha is 0.57 for NNF, 0.53 for NSF, 0.63 for NIH) (33). (iii) Third, we found that 88% of promotional words in the dictionary had their individual word percentage frequencies statistically correlated with the total frequency of all other promotional words ($p < 0.01$). To address concerns of the remaining 12% words, we ran separate analyses using only 88% of the promotional words and found that the results were not statistically different from our main finding, which uses all 139 words in the dictionary.

**Table 1**. Sample sentences from funded NIH grants showing examples of promotional words used in different contexts. Each promotional word is italicized and underscored (25).

| Example sentences using promotional words reported by Millar et al (2022) | NIH grant |
| --- | --- |
| "Further, a *unique* and *key* aspect of this program is the sharing of common mouse strains, reagents…" | R01AG032179 |
| "There remains an *imperative* need for more *advanced* PACT breast imaging technologies." | R35CA220436 |
| "Addressing this severe knowledge gap in one of the most *fundamental* aspects of cytoskeletal biology is *paramount* to understanding how actin functions in cells." | R35GM137959 |
| "The proposed methods offer a *revolutionary* innovation and will be a game-changer in the…" | R43EB027535 |
| "These *innovative* and *novel* studies will provide *essential* new information about the regulation of…" | R01HL084494 |
| "We propose to go deep in analyzing a very *unique* and *unprecedented* large scale human genomic data set for aging research." | R01AG055501 |

Our three outcome variables are the (i) funding success, (ii) innovativeness score, and (iii) future impact of the grant's publications. Funding success is coded as a binary variable (yes/no). Innovativeness score is measured using a widely used and validated novelty index that quantifies the degree to which a grant combines past knowledge in conventional or novel ways based on the historical referencing patterns (14, 34-39). Further computational details of this measure are provided in the Materials and Methods section. The future impact of papers based on the grant is quantified as the productivity and estimated citation impact of publications that acknowledge the funded grant. Following (40, 41), we estimate a grant's citation impact as (i) the average journal impact factor (JIF) and (ii) the largest JIF of publications acknowledging the grant. Productivity is measured as the number of publications supported by the grant.

Our predictor variable is operationalized as the percentage of promotional words in a grant proposal, i.e., "a grant proposal's total occurrences of promotional words" ÷ "a grant proposal's total number of words." We used regression models to test for the statistical correlations between promotional language and our outcome variables. Our regression models include control variables for PI characteristics (self-reported demographics, prior productivity and citation impact, and prior grant application and success), grant features (writing style, readability, length, applied funding amount), and fixed effects for year, grant type, and domain specializations (6, 17). S*I Appendix,* Table S1 presents variable definitions and operationalizations. *SI Appendix,* Section II presents the regression specifications, BIC, VIF, and cross-validation statistics, and robustness checks.

**Promotional Language and Grant Funding**

In our three datasets, the median percentage of promotional words per grant is one promotion word for every 100 words, or a density of one promotion word in every four sentences (an average sentence in the data contains 26 words). The average density of promotion words is higher in the first and last 500 words of a proposal, where first impression and recency biases tend to affect human recall and engagement the most (42).

Fig. 1 shows distributions of the percentage of promotional words in our datasets by funding decision. Comparing the two distributions of funded and unfunded grants indicates that funded grants contain statistically more promotional words than non-funded grants (p < 0.002 and p <0.01 for the t-test, KS-test, and Epps-Singleton non-parametric test in Fig. 1A and Fig. 1B, respectively).

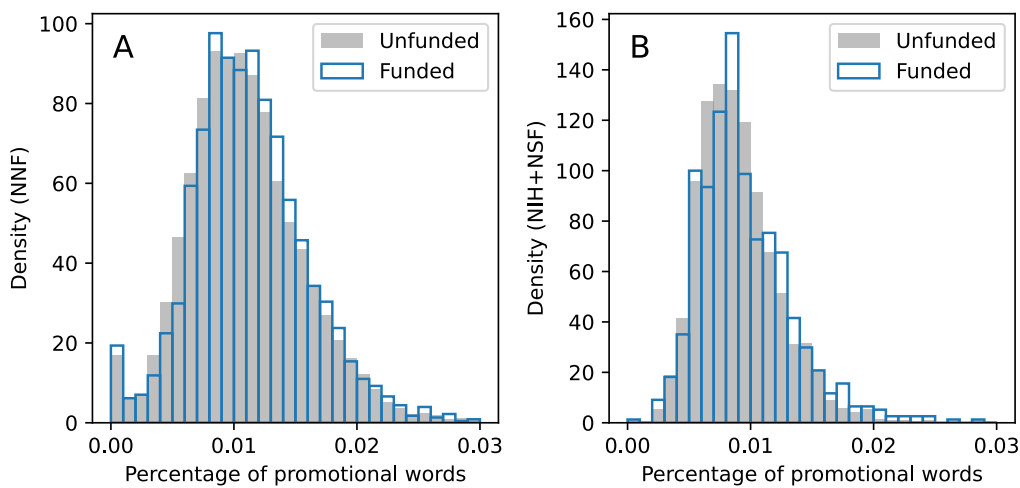

**Figure 1. Frequency Distribution of Promotional Language by Grant Funding Success.** Plots A and B show the frequency distribution of promotional words for NNF and NIH/NSF grants. On average, the density of promotional words per grant is one promotion word in every 100 words (1%) or one promotion word for every four sentences, with the density being highest in the first and last 500 words of a proposal, where promotional words appear once in every three sentences. Statistically, the plots indicate that funded NNF and NIH/NSF grants contain more promotional words

than do unfunded grants (p < 0.002 and p < 0.01 for the t-test, KS-test, and Epps-Singleton non-parametric test in both plots).

Logit regression demonstrates that promotional language predicts funding success. First, we regressed whether a NNF grant is funded or not on the percentage of promotional words in the grant while controlling for 13 variables representing a grant's semantic features, PI's prior productivity and citation impact, prior grant experience, self-reported gender and age, and fixed effects for application year, program area, and grant type (See *SI Appendix*, Table S1 for variable operationalizations and *SI Appendix*, Table S4 for regression coefficients).

Fig. 2A shows the margins plot from the logit regression for the NNF dataset. It indicates that increases in promotional language are significantly related to increases in the probability of a grant proposal being funded (β = 37.7, p < 0.001 in *SI Appendix*, Table S4). *SI Appendix*, Table S4 indicates that a 1 percentage point increase in the frequency of promotional words is associated with about 46% increase in the odds of a NNF grant being funded or a near doubling in funding probability, from a low of 11% to a high of 21%, after controlling for a grant's general semantic features and the PI's prior productivity and citation impact, prior grant experience, age, gender, and fixed effects for year, program area, and grant type. Fig. 2A also indicates that grants containing an average level of promotional words are funded below the 16.8% base level of success (dashed line). Specifically, the medium percentage of promotional words is 1.0% in the NNF dataset, which equates to an estimated acceptance rate of 14.8%, or two percentage points below the 16.8% base rate. Proposals that do better than the average probability of success contain 1.4 to 2.0 times the median percentage of promotional words (Fig. 2A).

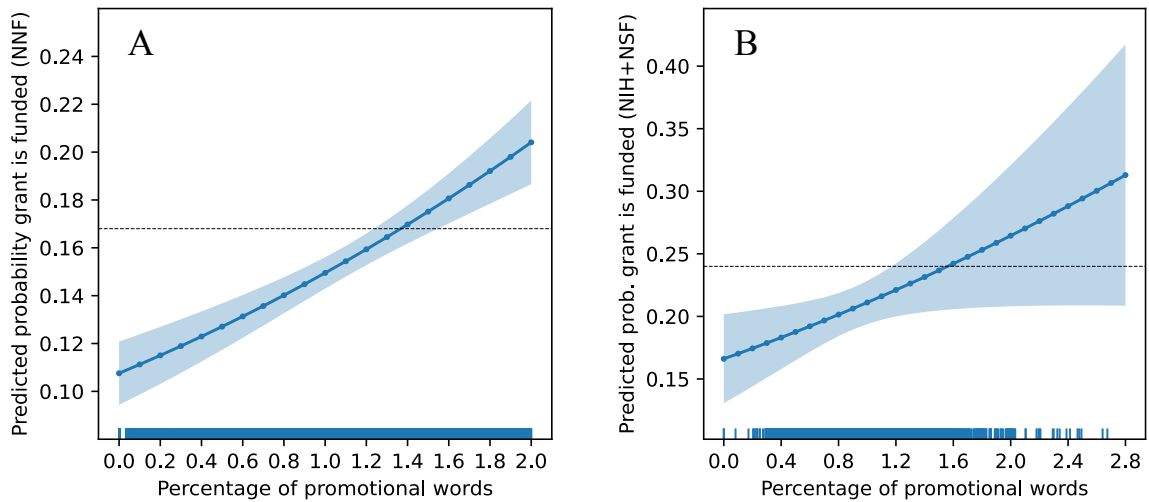

**Figure 2. Promotional Language Predicts Grant Funding Decision**. Margins plots with 95% CIs of the predicted probability of a grant proposal being funded as a function of the percentage of promotional words in the grant and control variables for a grant's general semantic features, the PI's prior productivity and citation impact, PI's prior grant experience, age, gender, and fixed effects for year, program area, and grant type. The two dashed lines represent the average acceptance rate of 16.8% in the NNF dataset (**A**) and 30.6% in the NSF/NIH combined dataset (**B**). The rug plots on the x-axis show the data's density distribution. In both datasets, the percentage of promotional words used

in a grant significantly predicts a positive funding decision ($p < 0.001$ in **A** and $p < 0.02$ in **B**) and its increase can double the likelihood of being awarded funding. In plot B, rerunning a regression with data only between 0.15% and 2.1% on the x-axis, which drops 26 proposals, produces a statistically significant result ($\beta = 28.3$, $p < 0.05$).

Fig. 2B shows the results for NIH/NSF grants ($\beta = 29.5$, $p < 0.02$ in *SI Appendix,* Table S6), which largely replicate the NNF finding. In the NIH/NSF dataset, the medium percentage of promotional words per grant is 0.9%, for which the model estimates an acceptance rate of 20.6%, or 3.4 percentage points below the 24.0% base rate of being funded. Thus, proposals with the median percentage of promotional words have an acceptance rate that is relatively 14.2% lower than the base rate. Proposals above the base acceptance rate have up to 3.0 times the median percentage of promotional words and a funding rate of up to 30% relative to proposals with the median number of promotional words. Given that there is data sparsity for promotional words above 2% in the NIH/NSF dataset as indicated by the rug plot, we reran a regression with data only in the range from 0.15% to 2.1%, which omits 26 observations in two tails of the distribution. The result indicates a slightly weaker but still statistically significant ($\beta = 28.3$, $p < 0.05$) link between promotional language and a positive funding decision in the NIH/NSF samples.

Finally, due to the rareness of data on failed grant applications in prior studies (6), we highlight our original findings on the associations between PI characteristics, grant features, and funding success. The regressions demonstrate that an applicant's prior number of citations and prior grant application success are positively correlated with being funded. By contrast, their number of prior publications and total number of prior grant applications are negatively linked to being funded. A proposal's general semantic features, including application length, reading score, and concreteness score are weakly related or unrelated to funding decision, which further highlights the unique semantic role of promotional language in funding evaluation (*SI Appendix,* Tables S4 and S6). Lastly, the estimated novelty of a NNF grant proposal is not statistically associated with funding acceptance (*SI Appendix*, Tables S5), corroborating existing evidence that grants tend to select for conventional ideas (11-13, 43).

**Promotional Language and a Grant's Inherent Innovativeness**

To examine whether promotional language reflects a grant's inherent level of innovativeness, we regressed a grant's innovativeness score on its percentage of promotional words while controlling for a grant's number of references (35, 39) and the 13 confounds included in the previous regression using the NNF dataset, which has the complete bibliography information needed to estimate a grant's innovativeness (35). Per our measure of innovativeness, less innovative grant applications combine prior knowledge in conventional ways, by referencing past work in statistically common ways based on the historical referencing patterns. By contrast, innovative proposals reference past work in statistically novel ways (14, 35, 36). The median innovativeness score in our data is 7.99 and the MAD (Median Absolute Deviation) is 6.13. In our NNF data, 1,970 of 13,520 applications lacked bibliographic data, reducing our observations in this analysis to 11,550. Statistical tests of the mean of all variables used in the regression showed that there were no statistical differences between the 11,550 samples and the full data. See Materials and Methods, and *SI Appendix*, Section IV for computational details of our innovativeness measure.

Fig. 3 shows the OLS regression's margins plot of the predicted innovativeness score as a function of a grant's percentage of promotional words while controlling for confounds. The plot indicates that promotional language positively and significantly correlates with a grant's innovativeness score ($\beta$ =139.7, p < 0.001; *SI Appendix,* Table S7 reports regression details). A proposal with the median percentage of promotional words of 1% has an estimated innovativeness score of 11.0, which is 38% higher than the median score in the data. Proposals with 2% of promotional words in them have an estimated innovativeness score of 12.3, which is 54% higher than the median score. Over the full range of estimated innovativeness scores, each one percentage point increase in promotional language is associated with a 1.4 (CI = [0.9, 1.9], p<0.001) point increase in the predicted innovativeness score.

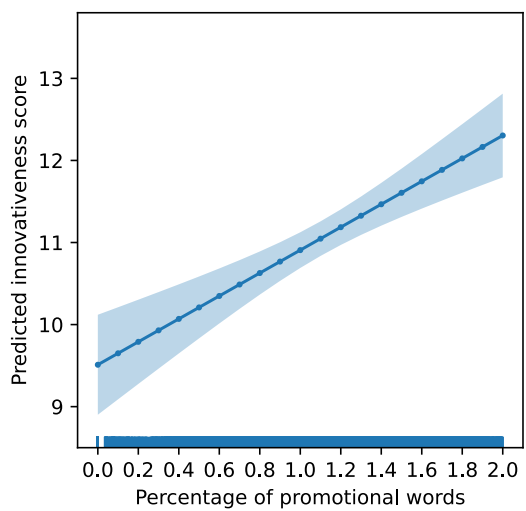

**Figure 3. Promotional Language is Proportional to a Grant's Innovativeness.** The margins plot from an OLS regression predicting a grant's innovativeness score based on its percentage of promotional language and controlling for confounds indicates that promotional language is a strong predictor of a proposal's innovativeness score (p < 0.001). On average, each one percentage point increase in promotional language corresponds to a 1.4-point (CI = [0.9, 1.9], p < 0.001) increase in innovativeness, which is equivalent to 17.5% of the median score.

**Promotional Language and Impact of Funded Grants**

While a paper's novelty and impact are positively related (34, 35), it is unknown whether a grant's promotional language is a leading indicator of the grant's citation and productivity impact. We use the same regression specification as in Fig. 2A, except we control for funding amount, we replace the variable "funding amount applied for" with "grant funded amount." We use OLS and Negative Binomial regressions to estimate the relationship between promotional language and a grant's (i) average journal impact factor (JIF), (ii) largest JIF, and (iii) number of papers based on grant-acknowledged publications. We used JIFs to estimate a paper's expected citations because recent papers lack the time needed to acquire citations (41) and because JIFs are good predictors of future citations when normalized by discipline (40). The grant-attributed publications are NNF-verified. At the grant level, the median average JIF is 5.7 (MAD = 2.0), the median highest JIF is 8.7 (MAD = 4.5), and the median productivity is 4.0 publications (MAD = 3.0). See *SI Appendix,* Tables S8-S11 for regression details.

Fig. 4A and 4B show that promotional language strongly predicts the expected average JIF ($\beta$ = 168.2, p < 0.001) and the highest JIF ($\beta$ = 366.6, p < 0.001) of a grant's papers (*SI Appendix*, Tables S8 and S9). A proposal with the median percentage of promotional language of 1% has an estimated average JIF of 6.9, a 21% increase over the median (Fig. 4A). Proposals that contain 2% of promotional words have an estimated average JIF of as high as 8.6, which is 51% higher than the median. When we examine a grant's max JIF, Fig. 4B shows that the promotion-impact association intensifies. For proposals with 2% of promotional words in them compared to proposals with 1% of promotional words in them, have an estimated increase in their max JIF by 30%, from 12.0 to 16.0. These results are robust (*SI Appendix*, Table S10) when using JIFs normalized by 14 academic disciplines defined in the UC San Diego map of science (44, 45). Finally, to address the concern that PIs have not yet reported recent publications, we run regressions on the subset of grants awarded before 2019 and find that the predictive power of promotional words for the avg/max JIF remains statistically significant (p < 0.001).

Promotional language also predicts productivity, but the link is statistically and substantively weaker ($\beta$ = 15.7, p < 0.02; *SI Appendix*, Tables S11) than citation impact. Fig. 4C shows the margins plot from a negative binomial regression predicting the number of publications reported by the PI while controlling for confounds. It indicates that a 1% increase in the frequency of promotional words is associated with about one more publication with or without controlling for the grant's estimated novelty (*SI Appendix*, Tables S12).

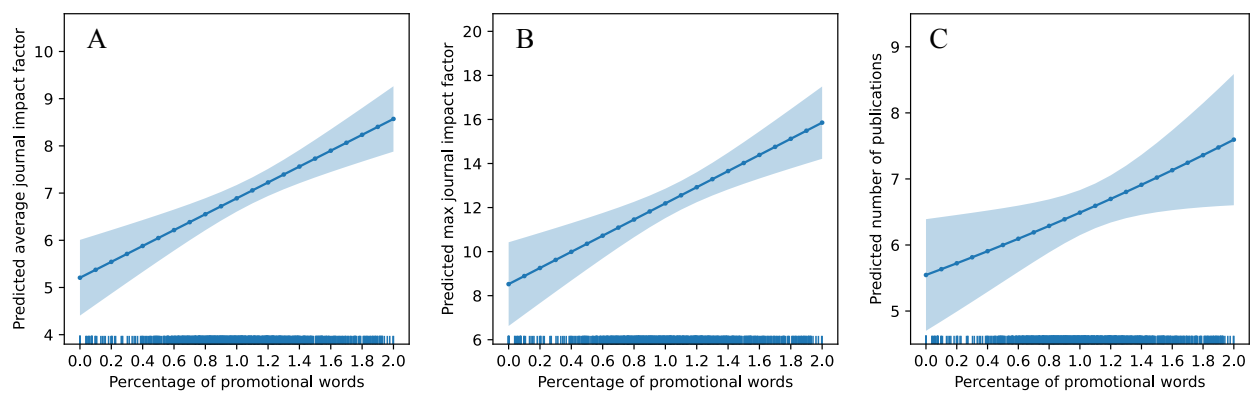

**Figure 4. Promotional Language Predicts Citation Impact and Productivity.** This figure shows margins plots from OLS (A and B) and Negative Binomial model (C) that regress the (**A**) average journal impact factor (JIF), (**B**) largest JIF, and (**C**) number of publications on the percentage of promotional language in the grant and controls. The frequency of promotional words strongly predicts the expected citation impact of publications (p < 0.001) and productivity (p < 0.02) of funded NNF grants.

## Robustness Checks on Statistical Analyses

Above, we discuss the additional psychometric validation tests we conduct on the Millar et al. promotional word lexicon. Here, we conduct three robustness checks to test the regression results sensitivity to measurement error in the use of the dictionary. First, to test the sensitivity of our analysis to the number of times a promotional word appears in a document, we run the analysis with each unique promotional word counted only once per grant proposal. Second, to test the context sensitivity of promotional words, e.g., the word "latest" can neutrally indicate a numerical ordering or promotionally indicate an idea's originality depending on the context in which it is used. Therefore, we ran analyses that randomly dropped up to 20% of a grant's promotional words. Third, to test our analysis's sensitivity to the misspecification of non-promotional words as promotional words, we run analyses with 5% to 20% of the promotional words in the dictionary randomly dropped from analyses. The above tests were run 100 times for the latter two conditions and indicated that the reported results are robust; while the significance level dropped as the level of measurement error increased, promotional words continued to predict funding at $p < 0.05$ (*SI Appendix*, Section I-B).

## Experimental Substitutions of Promotional Words and Estimated Changes in Sentiment

To examine how promotional language may help communicate the merits of innovative ideas, we conducted computer-assisted experiments that manipulated the promotional language in our grant samples. A starting point for our analysis is the potential link between promotional language and the perceived positive impressions of a grant's merits (46). Here we use algorithms to estimate the level of a grant's positive sentiment before and after promotional words are replaced with neutral synonyms, while leaving all other aspects of the text intact.

We collected each promotional word's set of synonyms from the Oxford Dictionary and employed a graduate student with two years of work experience in a university grant office to review and select neutral synonyms. For example, the word "necessary" is synonymous with the promotional word "critical" but has a neutral connotation according to human raters (25). The manual review process also ensured that each synonym is valid within a scientific context (e.g., the word "handsome" is a valid synonym for the promotional word "attractive" but is out of the grant writing context). See details and the full list of non-promotional synonyms in *SI Appendix,* Section V. Each of the 139 promotional words has an average of 13 neutral synonyms.

**Table 2**. Example sentences containing promotional words and their substitutions with randomly chosen neutral synonyms.

| Example sentence with promotional words | Non-promotional replica with replacement |
|---|---|
| "Further, a *unique* and *key* aspect of this program is the sharing of common mouse strains, reagents…" | "Further, a *specific* and *central* aspect of this program is the sharing of common mouse strains, reagents…" |
| "There remains an *imperative* need for more *advanced* PACT breast imaging technologies." | "There remains a *necessary* need for more *modern* PACT breast imaging technologies." |
| "The proposed methods offer a *revolutionary* innovation and will be a game-changer in the…" | "The proposed methods offer a *different* innovation and will be a game-changer in the…" |

We used *SciBert* to estimate a grant's positive sentiment, which is a contextual word embedding model pre-trained on scientific publications and can be used to evaluate the sentence-level sentiment of scientific text (47). The model outputs a confidence score for each of three sentiment labels (positive, neutral, negative). In the experiment, we focus on positive sentiment score because most promotional words contain positive emotion. To fine-tune *SciBert* for the study of sentiment in grant applications, we followed a standard fine-tuning methodology (48) by further training *SciBert* on sentences from grants with labeled sentiment. We randomly sampled 1,021 sentences from 40 NSF and 40 NIH abstracts published in 2024. These sentences were given to three raters who manually coded the sentences as having positive, neutral, or negative sentiment. Rater 1 was a native English speaker and a psychology graduate with two years of work experience reviewing grant submissions at a large university. Rater 2 was a doctoral candidate who speaks English as a second language and has two years of research experience. Both raters were unaware of the purpose of our substitution experiment. Rater 3 was one of the authors. The three raters worked independently. Pairwise, raters agreed on 66%, 76%, and 67.5% of all sentences, and Cohen's kappa is 0.29, 0.53, and 0.38, which is considered acceptable agreement. A sentence was labeled as the category on which two or more of the raters agreed. In the handful of cases that lacked majority agreement, Rater 1 and Rater 2 resolved the sentence's coding.

The fine-tuning procedure started with 200 sentences and incremented the sample by 50 sentences at a time up to 1,020. For each sample size, we fine-tuned *SciBert* five times, each time using 85% of the data as training and 15% as testing. In the testing set, *SciBert* reached an F1 score of 0.79 when using the entire dataset. This fine-tuning process indicates that when using more than 500 sentences, the F1 score stabilizes (See *SI Appendix*, Figure 3).

For each promotional-word-containing sentence in each grant, we replaced the promotional word with a randomly chosen neutral synonym. We then compared whether the replicant sentences' average positive sentiment went up, went down, or stayed the same with respect to the original sentiment. A trial was coded as a decrease in positive sentiment if the average positive sentiment goes down. This process was repeated 100 times for each proposal to remove random noise and outlier effects. To assess whether the replicant proposal has a statistically lower sentiment score than the original proposal, we used a binomial test (N = 100 trials, K = the number of trials the original proposal had a higher average positive sentiment than its substitution versions). This entire procedure was run for independent computer experiments that replaced 25, 50, 75, and 100 percent of a proposal's promotional words. In total, these experiments were run on 13,520 NNF, 2,649 NIH, and 561 NSF grant proposals.

Fig. 5 shows the percentage of proposals with a statistically higher level of positive sentiment than their neutral replicas at four levels of substitutions based on the binomial test. Across all three datasets, a super-majority of original proposals have significantly more positive sentiment than their replicas, holding the meaning and content of the proposal constant. This positive sentiment premium intensifies when the level of word substitutions increases. In the NSF and NIH datasets, over 69% and 68% (respectively) of proposals show a sentiment drop when 25% of promotional words are substituted and over 88% and 92% of proposals show a sentiment decrease when 100% of promotional words are substituted. The NNF dataset shows a similar trend. These results suggest that promotional words are linked with higher estimated positive impression of a grant.

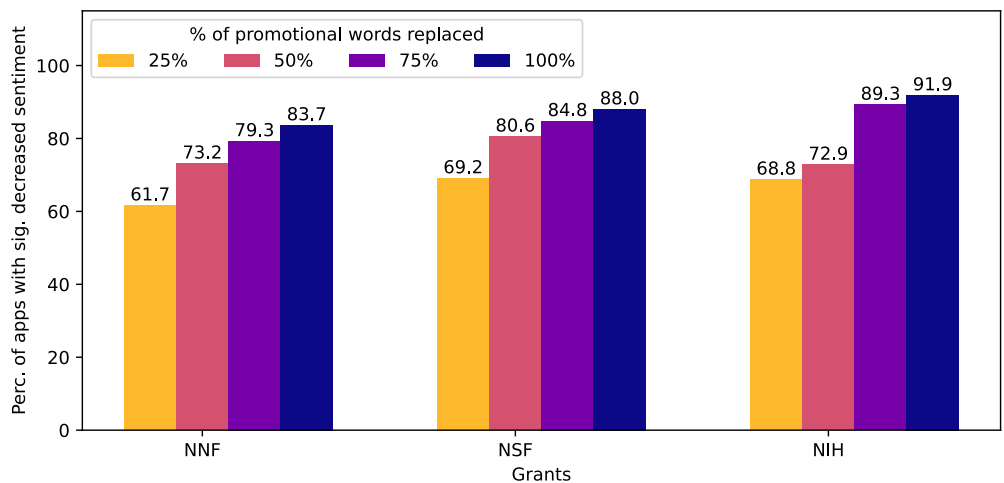

**Figure 5. A grant proposal's positive sentiment drops when its promotional language is experimentally replaced with neutral synonyms**. The y-axis shows the percentage of grant applications that decrease in positive sentiment after substituting a proposal's observed promotional words with their non-promotional synonyms chosen at random. Significance tests are within-proposal and based on 100 rounds of random substitutions for 13,520 NNF, 2,649 NIH, and 561 NSF grant proposals. At all levels of replacement, the estimated sentiment score of a proposal drops when promotional words are replaced with synonyms that lack a promotional connotation as coded by human raters.

A robustness analysis confirmed the expectation that a proposal's average sentiment score is positively associated with funding acceptance. Specifically, we used our fine-tuned *SciBert* model to estimate the average sentence-level sentiment of each NNF grant and ran three regressions. Regression one regressed funding success on a grant's average sentiment plus basic controls for year, grant type, and program area. Consistent with expectations, it showed that average grant sentiment predicts funding (p<0.001, BIC=12,025). Next, we ran a second regression with the same specification as the first one, except we replaced average grant sentiment with the percentage of promotional words. Regression two indicated that the percentage of promotional words predicts funding (p<0.001, BIC=12,007). To evaluate the covarying effects of average sentiment and the percentage of promotional words, regression three regressed funding on both average grant sentiment and percentage of promotional words. Regression three indicated that promotional language was statistically significant (p<0.001) and the average grant sentiment was statistically insignificant (p>0.05, BIC = 12,013). The fact that the "promotional word only" regression (regression two) had the lowest BIC of all three regressions provides further evidence that the percentage of promotional words is the stronger predictor of funding(49). Similarly, this robustness test replicates (see details in *SI Appendix*, Table S13) when using the full regression model presented in *SI Appendix*, Table S4.

## Discussion

Innovation is a distinctive strength of science. In science, researchers have accumulated much knowledge about conditions that generate innovative ideas, such as teams, networks, institutional support, training, incentives, and diversity (4, 36, 50, 51). Comparatively, little is known about how meritorious ideas are communicated to other scientists and the entities that support science (35, 52). Indeed, a lack of effective communication and factors associated with communication skills may partly explain the observed increase in career setbacks (6, 14), inequality in science (17, 53-55), and resistance to scientific facts (56).

Our starting point for examining the communication of scientific ideas focuses on the role of promotional language in grant applications. Adopting a semantic analysis approach, we examined the link between promotional language and grant funding success. The density of promotional language has skyrocketed in grants and on social media (25, 55), yet claims of its role in science remain controversial. On the one hand, promotional language is thought to increase attention to, and connections between, concepts in ways that help convey the potential of new ideas. On the other hand, it may result in misleading levels of hype that turn off reviewers or hurt science's reputation for conservatively presenting findings (28). Our study used a promotional language lexicon identified and validated by Millar et al. (25) to examine the relationship between the density of promotional language in a grant proposal and a grants' (i) probability of funding, (ii) degree of innovativeness, and (iii) productivity and citation impact in the published literature.

Our analysis had several features that can help to mitigate weaknesses in previous research on grants. First, we analyzed tens of thousands of grant applications from three diverse funding organizations – the public NIH, NSF, and the private Novo Nordisk Foundation, the world's largest philanthropic foundation by endowment. Second, to surmount sample selection biases found in prior work, our data includes a proposal's full text, all funded and unfunded proposals, and longitudinal measures of PI, grant, and agency characteristics over the last 10 years to control for confounds.

Our analyses showed that across diverse datasets, the percentage of promotional language in a grant, net of key confounds, reliably predicts a grant's funding decision, innovativeness score, and productivity and citation impact. Our raw data indicates that on average a grant contains one promotional word in every 100 words, or about one promotional word every fourth sentence. Adjusting for confounds, proposals with above average probability of funding success contain 1.4 to 2.0 times the average percentage of promotional words, which at the limit of our data can more than double a grant's chances of funding.

To understand the mechanisms through which promotional language might shape impressions about a grant proposal, we conducted word substitution experiments to estimate the level of sentiment produced by promotional words. To predict a grant's level of positive sentiment, these experiments used *SciBert*, a domain-specific and context-sensitive foundational word embedding model in the context of scientific writing, along with fine-tuning based on over a thousand sentences from grants that were manually coded for their positive and negative sentiment. By comparing a proposal's positive sentiment before and after the substitution of promotional words with neutral synonyms, ceteris paribus, (e.g., substitute the promotional word "unique" with a neutral synonym "specific"), we found that the replacement of promotional words with their neutral synonyms results significant drop in a grant's positive sentiment. While the algorithm cannot capture all nuances in academic language and phrasing, and are only be a surrogate for how humans might respond to promotional language, the experimental results agree with theory (20) that promotional language is associated with reviewers' subjective impressions (57, 58). We hope that future research will use our initial analysis to conduct direct causal tests of this mechanism and others. A related question for future research is why promotional language has increased and how it propagates through networks in science (51).

Our findings raise important implications for scientific practice. Granting agencies have long urged that science be communicated in plain language (23). Our findings suggest a need to understand the gap between science's espoused norm of neutrality and the practical need for effectively communicating meritorious ideas and findings. On the one hand, the conservative approach of science, which embraces the ideal that good ideas are recognized based on their face value, has served it well. Yet, the astonishing growth of scientific knowledge and changes in search technology and the proliferation of subfield journals have led to the observation that reviewers' attention is becoming a scarce commodity – changes that can make the recognition of innovation harder and more parochial (35, 59-62) and that additionally burdens researchers to convey their ideas to multiple audiences who hold different perspectives on innovativeness (30, 34, 44). From a theory and policy perspective, promotional words, and semantic analysis more generally, can help increase scientists' knowledge of how to connect and draw attention to boundary-spanning ideas. Mindful of these cross-cutting needs, one speculation for policy analysis is to examine how promotional language can help bridge the gap between recognizing innovative ideas and the changing reality of reviewer's attention and specialization in science.

Another policy issue is whether knowledge of the association between promotional language and grant success can inadvertently create hype. We believe that science is built on trust and that the pervasive trustworthiness scientists show in data integrity and honesty in reporting would apply to the use of promotional language as well. Consistent with this belief, our results showed that the percentage of promotional words in a grant statistically correlates with the grant's intrinsic level of innovativeness. To better understand the use of promotional language and norms in science, future research might begin to investigate how promotional language varies with changing norms like data-sharing, replicability, and social media discourse.

A scope condition of our research is that we examined grants specifically. Papers and patents are also critical to academic success and to the conversion of scientific ideas into facts and inventions (63). It may be that the findings for grants are different in the context of papers or patents. Grants are designed to gain funding to support aspirational research ideas, which may invite and justify the use of more promotional language than is used in papers and patents, which are geared toward presenting proof of empirical findings and concepts. Thus, while grants are uniquely important to science, the findings here may be grant-specific. It remains an open question as to whether promotional language has positive, negative, or non-linear associations within the context of papers and patents. Consequently, an important next step is to examine how promotional language may generalize to academic papers and patents, a question we will investigate in follow-up work.

Beyond promotional language, our study speaks to the growing interest in how semantic analysis of science can improve training and accelerate discovery (19). While science rightly focuses on the quantification of phenomena, most of the content of a grant application, scientific paper, or patent application is text. Most text has been an undeveloped research asset until recently. Our work offers a novel demonstration of a growing research area (20) on how data analytics, linguistic theory, NLP, and machine learning technologies can provide intriguing approaches for advancing science and its contributions to society.

**Materials and Methods**

**Promotional Words Lexicon and Validation**. We operationalized promotional words as a list of 139 hype words identified in Millar et al. 2022 (25). Millar et al. examined 901,717 NIH-funded grants from 1985-2020 and created a funding-specific lexicon of promotional language in NIH grant writing (25). In Millar et al.'s (25) sophisticated manual coding process, an adjective was labeled as a promotional word based on whether the word could be removed or replaced with a more objective or neutral word without changing the information within the sentence. Each candidate word was reviewed in at least 500 different instances of its use. In their hand-curation process, independent experts achieved a strong inter-rater agreement (Cohen's $k$=0.82) and resolved disagreement through discussion (25).

In addition to the human validations performed by Millar et al., we conducted three tests to check the validity of the promotional word dictionary, and three robustness checks to test the sensitivity of our main finding to measurement errors. Both analyses are described in the main text with further details reported in the *SI Appendix*, Section I.

The lexicon of 139 scientific promotional words includes: compelling, critical, crucial, essential, foundational, fundamental, imperative, important, indispensable, invaluable, key, major, paramount, pivotal, significant, strategic, timely, ultimate, urgent, vital, creative, emerging, first, groundbreaking, innovative, latest, novel, revolutionary, unique, unparalleled, and unprecedented, accurate, advanced, careful, cohesive, detailed, nuanced, powerful, quality, reproducible, rigorous, robust, scientific, sophisticated, strong, systematic, accessible, actionable, deployable, durable, easy, effective, efficacious, efficient, generalizable, ideal, impactful, intuitive, meaningful, productive, ready, relevant, rich, safer, scalable, seamless, sustainable, synergistic, tailored, tangible, transformative, user-friendly, ambitious, collegial, dedicated, exceptional, experienced, intellectual, longstanding, motivated, premier, prestigious, promising, qualified, renowned, senior, skilled, stellar, successful, talented, vibrant, ample, biggest, broad, comprehensive, considerable, deeper, diverse, enormous, expansive, extensive, fastest, greatest, huge, immediate, immense, interdisciplinary, international, interprofessional, largest, massive, multidisciplinary, myriad, overwhelming, substantial, top, transdisciplinary, tremendous, vast, attractive, confident, exciting, incredible, interesting, intriguing, notable, outstanding, remarkable, surprising, alarming, daunting, desperate, devastating, dire, dismal, elusive, stark, unanswered, unmet. See *SI Appendix*, Figure S1 for the frequency of promotional words in our datasets.

**Grant Innovativeness Measure.** We estimated a grant's innovativeness using a popular and validated measure that characterizes a grant's novelty based on the degree to which it combines past knowledge in either familiar or novel ways (35, 36). The ideas combined in a grant are reflected in the references co-cited in its bibliography. Bibliographies that contain references that have been paired together many times in prior work combine ideas in conventional and familiar ways. References that have never or rarely been paired together in prior work combine ideas in novel and innovative ways.

Journal pairs in a paper's bibliography are used to measure the pairing of knowledge. To quantify the novelty of a given journal pair, the method computes an observed co-citation frequency based on the pairs observed in past published work and compares it with the frequency expected by chance based on a null model that shuffles citations while preserving the yearly citation statistics. The preceding procedure creates a z-score for each journal pair and a distribution of z-scores for each grant (one score for each journal pairing in its bibliography). Z-scores below zero statistically happen less than chance and vice versa for z-scores above zero. Thus, scores above zero are increasingly conventional and scores below zero are increasingly innovative. A proposal's innovativeness score is defined as the median of its negative scores. The innovativeness score was reverse coded so that large scores indicate more innovativeness. *SI Appendix*, Section IV presents the procedure's full details and an example.

**Data Availability.** The deidentified NNF grant application dataset is available upon request through an NDA process with the Novo Nordisk Foundation (NNF). The NSF and NIH datasets used in our analysis are subject to their University's NDA process. This study is approved by Northwestern University's IRB office (STU00219074 & STU00215754).

**ACKNOWLEDGMENTS.** We thank the Novo Nordisk Foundation, Northwestern University, Clarivate/Web of Science and OpenAlex for providing the data used in this study, especially Anders A.K. Nielsen (NNF), Northwestern's High-Performance Computing Cluster for technical contribution to this study, Emily Rosman and Zihang Lin's research assistance, and the Center for Science of Science and Innovation workgroup at Northwestern University for feedback on earlier drafts. This work is funded by the Air Force Office of Scientific Research under Minerva award number FA9550-19-1-0354, the Northwestern Institution on Complex Systems and Data Analytics (NICO), and the Kellogg School of Management at Northwestern University.